\documentclass[letter]{ptptex}

\usepackage{graphicx}
\usepackage{color}

\def\non{\nonumber}

\def\bfm#1{\mbox{\boldmath $#1$}}

\def\d2der#1#2#3{\frac{d^2 #1}{d #2 d #3}}

\def\p2der#1#2#3{\frac{\partial^2 #1}{\partial #2 \partial #3}}
\def\bra#1{\langle#1|}
\def\ket#1{|#1\rangle}

\def\fun#1#2{\lower3.6pt\vbox{\baselineskip0pt\lineskip.9pt
\ialign{$\mathsurround=0pt#1\hfil##\hfil$\crcr#2\crcr\sim\crcr}}}

\notypesetlogo                       

\title{
New Approach for Evaluating Incomplete and Complete Fusion Cross Sections
with Continuum-Discretized Coupled-Channels Method
}

\author{
Shintaro \textsc{Hashimoto}$^{1,}$\thanks{E-mail: hashimoto.shintaro@jaea.go.jp},
Kazuyuki \textsc{Ogata}$^{1,2}$,
Satoshi \textsc{Chiba}$^{1,3}$, and \\
Masanobu \textsc{Yahiro}$^{2}$
}

\inst{
$^1$Advanced Science Research Center, Japan Atomic Energy Agency, \\Ibaraki 319-1195, Japan
\\
$^2$Department of Physics, Kyushu University, Fukuoka 812-8581, Japan
\\
$^3$National Astronomical Observatory of Japan, Tokyo 181-8588, Japan
}

\abst{
We propose a new method for evaluating incomplete and complete fusion cross
sections separately using the Continuum-Discretized
Coupled-Channels method.
This method is applied to analysis of the deuteron
induced reaction on a $^7$Li target up to 50 MeV of the deuteron
incident energy.
Effects of deuteron breakup on this reaction are explicitly taken into
account.
Results of the method are compared with those of the Glauber model,
and the difference between the two is discussed.
It is found that the energy dependence of the incomplete fusion
cross sections obtained by the present calculation is almost the
same as that obtained by the Glauber model, while for the
complete fusion cross section, the two models give markedly
different energy dependence.
We show also that a prescription for evaluating incomplete fusion
cross sections proposed in a previous study gives much smaller
result than an experimental value.
}

\begin{document}

\maketitle

The understanding of the fusion reaction mechanism is one of the most
important and challenging subjects of nuclear physics.
Description of {\it incomplete} fusion processes,
in which a part of the projectile is absorbed by the target nucleus,
with emitting other projectile fragment(s),
is particularly interesting and important.
So far, some theoretical models of the incomplete fusion, also
called breakup fusion or inclusive breakup, of a two-body projectile
have been proposed.\cite{UT,IAV,HM} \
In these models, the incomplete fusion reaction was described as
two-step processes, i.e., the projectile is broken up first and then
one of the two constituents is absorbed by the target.
The calculations of the fusion cross sections were carried out by
using the Distorted Wave Born Approximation (DWBA) assuming that
the emitted fragment can be treated as a spectator in the
final state.
Recently, roles of breakup (continuum) states of a weakly-bound
projectile in the fusion reaction have been
discussed \cite{Diaz1,Diaz2} using the Continuum-Discretized
Coupled-Channels method (CDCC).\cite{CDCC} \
CDCC was proposed and developed by Kyushu group, and has been
successfully applied to analyze various reaction processes;
see, e.g., Refs.~\citen{Ogata,FBCDCC,Surrey}.
In Ref.~\citen{Diaz1}, an attempt to calculate the incomplete
and complete fusion cross sections separately with CDCC was
described. The assumption used in the separation of the two was,
however, unrealistic for some reasons; we will return to this point
later.

The incomplete fusion process in a deuteron induced reaction
on Li targets at incident energies up to 50 MeV attracts
wide interests of not only nuclear physicists but also
nuclear engineers, because
the emitted neutrons through this process are planning to be used
in the International Fusion Materials Irradiation Facility
(IFMIF).\cite{Matsui} \
Understanding of the reaction mechanism of this incomplete fusion
process, or, equivalently, the inclusive $(d,n)$ process, with
evaluating theoretically the absolute value of the cross section
is necessary. Moreover, nuclear data of inclusive $(d,n)$ reactions
on other various targets such as Be, Ta, and W are of crucial
importance for studies on accelerator-based applications,
i.e., shielding of the deuteron accelerators including IFMIF,
and medical applications for Boron Neutron Capture Therapy (BNCT).
Very recently, Ye {\it et al.} \cite{Ye3} showed that the
main part of the double differential cross section (DDX)
data \cite{Hagiwara} of the
emitted neutron from the inclusive $^7$Li$(d,n)$ reaction at 40 MeV
is reproduced very well by
the proton stripping cross section
$d^2 \sigma_{\rm STR}^{(p)}/(d E_n d \Omega_n)$,
which corresponds
to the ($d,n$) incomplete fusion process in our terminology, added by
the elastic breakup cross section $d^2 \sigma_{\rm EB}/(d E_n d \Omega_n)$;
$E_n$ and $\Omega_n$ are the energy and solid angle of the outgoing neutron.
In their study, $d^2 \sigma_{\rm STR}^{(p)}/(d E_n d \Omega_n)$
and $d^2 \sigma_{\rm EB}/(d E_n d \Omega_n)$ are obtained by
the Glauber model and CDCC, respectively,
and the reason
for the surprising success of the Glauber model at such low energies
($\sim 40$ MeV) was discussed.\cite{Ye3} \
Nevertheless, it is important to evaluate the accuracy of the
Glauber model calculation of $d^2 \sigma_{\rm STR}^{(p)}/(d E_n d \Omega_n)$
below 40 MeV, in which experimental data are very scarce,
by comparing it with that obtained fully quantum mechanically.

In this Letter, we propose a new approach for calculating
the complete and incomplete fusion cross sections separately using CDCC.
As an important advantage of the present method to the preceding
studies,\cite{Diaz1,Iijima} we separate the two fusion processes by the
physics condition on the absorption of each constituent of the projectile
by the target nucleus. In our model, a possible contribution of the
breakup channels to the complete fusion process,
as well as that of the elastic channel
to the incomplete fusion process
is properly taken into account.
As we mention below, the new method contains a free parameter,
i.e., the absorption radius. This parameter is determined using
the result of the Glauber calculation at 40 MeV that
can be interpreted as an experimental value of the proton-stripping
incomplete fusion cross section.
We then apply this method to the $^7$Li$(d,n)$ reactions at
different deuteron incident energies from 10 MeV to 50 MeV.
The $(d,n)$ and $(d,p)$ incomplete fusion cross sections and the
complete fusion cross section thus evaluated are compared with
the results \cite{Ye3,Watanabe} obtained by the Glauber model.
Note that we focus on the cross sections integrated over emission
energies and angles in this work.

\begin{figure}[!t]
\centerline{\includegraphics[width=50mm,keepaspectratio]{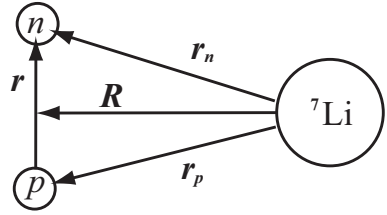}}
\caption{\label{fig-System}
Illustration of the three-body model of $p$, $n$ and $^7$Li.
}
\end{figure}

We describe the $^7$Li$(d,n)$ reaction with the three-body system shown
in Fig.~\ref{fig-System};
$\bfm{R}$ is the relative coordinate between the $^7$Li target
and the center of mass of $d$, and $\bfm{r}$ is the relative
coordinate between $p$ and $n$.
The coordinate of $p$ ($n$) relative to $^7$Li is denoted by
$\bfm{r}_p$ ($\bfm{r}_n$). The three-body Hamiltonian is given by
\begin{eqnarray}
 H =
  T_R + U_p(r_p) + U_n(r_n) + V^{\rm Coul}(R) + T_r + V_{pn}(r),
\label{eq-Hamiltonian}
\end{eqnarray}
where $T_R$ and $T_r$ represent the kinetic energy operators
associated with $\bfm{R}$ and $\bfm{r}$, respectively,
$U_p$ ($U_n$) is the optical potential between $^7$Li and $p$ ($n$),
$V^{\rm Coul}$ is the Coulomb interaction between $d$ and $^7$Li,
and $V_{pn}$ is the interaction between $p$ and $n$.
Note that we neglect the Coulomb breakup processes in this study,
since we are interested in the ($d,n$) reactions enough above the Coulomb
barrier energy.

In CDCC, the three-body wave function $\Psi(\bfm{R},\bfm{r})$ is
expanded in terms of the eigenfunctions of the $p$-$n$ system
$\{ \hat{\Phi}_i(\bfm{r}),\; i=0$--$i_{\rm max} \}$:
\begin{eqnarray}
 \Psi(\bfm{R},\bfm{r}) =
  \sum_{JM} \sum_{i=0}^{i_{\rm max}}
  \left[\chi_i(\bfm{R}) \otimes \hat{\Phi}_i(\bfm{r})\right]_{JM},
\label{eq-WF}
\end{eqnarray}
where $\hat{\Phi}_0$ represents the ground state of $d$
and $\hat{\Phi}_i$ ($i\ne 0$) the $i$th discretized continuum state.
The {\it expansion coefficient} $\chi_i(\bfm{R})$ describes the relative
motion between $d$ in the $i$th state and $^7$Li.
The discretized continuum state $\hat{\Phi}_i$ is obtained by
the so-called average method \cite{CDCC} as
\begin{eqnarray}
 \hat{\Phi}_i(\bfm{r}) =
  \frac{1}{\sqrt{\Delta k_i}} \int_{k_{i-1}}^{k_i} dk
  \Phi(\bfm{r},k),
\label{eq-discretization}
\end{eqnarray}
where $\Phi(\bfm{r},k)$ is the $p$-$n$ scattering wave function
with the relative wave number $k$, and $\Delta k_i=k_i-k_{i-1}$.
$\Phi(\bfm{r},k)$ satisfies
\begin{eqnarray}
 \left[T_r + V_{pn}(r)\right] \Phi(\bfm{r},k) =
  \varepsilon \Phi(\bfm{r},k)
\label{eq-ISchrodinger},
\end{eqnarray}
where $\varepsilon=\hbar^2 k^2 /(2 \mu_r)$ with $\mu_r$
being the reduced mass of $p$ and $n$.

The three-body Schr\"odinger equation using the wave function
of Eq.~(\ref{eq-WF})
is given by
\begin{eqnarray}
(H-E)\Psi(\bfm{R},\bfm{r})=0,
\label{eq-IISchrodinger}
\end{eqnarray}
where $E$ is the total energy. Multiplying Eq.~(\ref{eq-IISchrodinger})
by $\hat{\Phi}_j^*$ from the left,
and integrating over $\bfm{r}$, we obtain the following
coupled-channel equations for $\chi_i(\bfm{R})$:
\begin{eqnarray}
 \left(T_R + V_p^{\rm Coul}(R) + \varepsilon_i - E\right) \chi_i(\bfm{R}) =
  -\sum_j F_{ji}(\bfm{R}) \chi_j(\bfm{R}),
\label{eq-CDCC}
\end{eqnarray}
where $\varepsilon_i$ is the internal energy of the $p$-$n$ system in the
$i$th state and
\[
F_{ji}(\bfm{R})\equiv\bra{\hat{\Phi}_j}(U_p+U_n)
\ket{\hat{\Phi}_i}_{\bfm{r}}
\]
is the coupling form factor.
Equations (\ref{eq-CDCC}) are solved under the usual
boundary conditions for $\chi_i(\bfm{R})$.~\cite{CDCC}

The imaginary part of the optical potential is considered to
describe the particle absorption by the target nucleus.
Thus, the fusion cross section (absorption cross section) is given
as the expectation value of the imaginary part with the wave function
of the system. This cross section contains both contributions from
the complete and incomplete fusion processes;
we henceforth call this the total fusion cross section $\sigma_{\rm TF}$.
In the present three-body model calculation,
$\sigma_{\rm TF}$ is obtained by
\begin{eqnarray}
 \sigma_{\rm TF} =
  \frac{2\mu_R}{\hbar^2K_0}
  \left|\langle \Psi|(-W_p-W_n)|\Psi \rangle\right|,
\label{eq-TFX}
\end{eqnarray}
where $W_p$ ($W_n$) is the imaginary part of $U_p$ ($U_n$),
$\mu_R$ is the reduced mass between $d$ and $^7$Li, and $K_0$
is the $d$-$^7$Li relative wave number in the incident channel.
Note that the integrand on the right hand side (r.h.s.)
of Eq.~(\ref{eq-TFX}) is compact ($L^2$ integrable),
since we discretize the $p$-$n$ scattering wave functions with
Eq.~(\ref{eq-discretization}).
Another important point to be noted is that the imaginary
part of the nucleon-$^7$Li optical potential describes not only
nucleon absorption but also other processes such as the
inelastic scattering to the excited states of $^7$Li.
Since the nucleon absorption
has the main contribution to the r.h.s. of Eq.~(\ref{eq-TFX}),
however, we regard it as the total {\lq\lq}fusion'' cross section as in many
other studies on fusion reactions.

\begin{figure}[!t]
\centerline{\includegraphics[width=60mm,keepaspectratio]{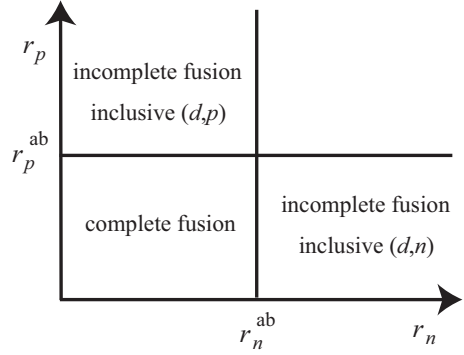}}
\caption{\label{fig-Integ}
Schematic illustration of the four integration regions.
See text for details.
}
\end{figure}

To separate the $(d,p)$ and $(d,n)$ incomplete fusion cross sections,
$\sigma_{\rm IF}^{(n)}$ and $\sigma_{\rm IF}^{(p)}$ respectively, from
$\sigma_{\rm TF}$, we divide the integration region in Eq.~(\ref{eq-TFX})
as follows. The explicit form  of the expectation
value on the r.h.s. of Eq.~(\ref{eq-TFX}) is given by
\begin{eqnarray}
 \left|\langle \Psi|(-W_p-W_n)|\Psi \rangle\right| =
  -\int d\bfm{r}_p \int d\bfm{r}_n \Psi^*(\bfm{R},\bfm{r})
  \{ W_p(\bfm{r}_p)+W_n(\bfm{r}_n) \} \Psi(\bfm{R},\bfm{r}),
\label{eq-TFX2}
\end{eqnarray}
where we have changed the integration variables from
($\bfm{R}$, $\bfm{r}$) to ($\bfm{r}_p$, $\bfm{r}_n$).
We now separate the integration regions over $r_p$ and $r_n$ as
\begin{eqnarray}
 \int d\bfm{r}_p \int d\bfm{r}_n
  &=&
  \int_{r_p<r_p^{\rm ab}}  d\bfm{r}_p
  \int_{r_n<r_n^{\rm ab}}  d\bfm{r}_n
  +\int_{r_p<r_p^{\rm ab}}  d\bfm{r}_p
  \int_{r_n>r_n^{\rm ab}}  d\bfm{r}_n
  \non \\
 &&
  +\int_{r_p>r_p^{\rm ab}}  d\bfm{r}_p
  \int_{r_n<r_n^{\rm ab}}  d\bfm{r}_n
  +\int_{r_p>r_p^{\rm ab}}  d\bfm{r}_p
  \int_{r_n>r_n^{\rm ab}}  d\bfm{r}_n,
\label{eq-Interval}
\end{eqnarray}
where $r_c^{\rm ab}$ ($c=p$ or $n$) is the interaction range of $W_c$;
at $r_c > r_c^{\rm ab}$, $W_c$ is assumed to be negligible.
The first term on the r.h.s. of Eq.~(\ref{eq-Interval}) corresponds
to the process in which both $p$ and $n$ are located within the
range of $W_c$ and absorbed by $^7$Li.
In the second term, $p$ is assumed to be within
the range of the absorbing potential, while $n$ is free of the absorption.
Thus, it gives the integration region corresponding to the ($d,n$) incomplete
fusion process. Similarly, the third term corresponds to the ($d,p$)
incomplete fusion process.
It is obvious from the definition of $r_c^{\rm ab}$
that the fourth term gives no contribution to $\sigma_{\rm TF}$.
Schematic illustration of these four integration regions is shown in
Fig.~\ref{fig-Integ}.
Using Eq.~(\ref{eq-Interval}), $\sigma_{\rm TF}$ is decomposed into
the complete fusion cross section $\sigma_{\rm CF}$ and
the above-mentioned two incomplete fusion cross sections, i.e.,
\begin{eqnarray}
 \sigma_{\rm TF} =
 \sigma_{\rm CF} + \sigma_{\rm IF}^{(p)} + \sigma_{\rm IF}^{(n)},
\label{eq-TFX3}
\end{eqnarray}
where
\begin{eqnarray}
\sigma_{\rm CF}=
  \frac{-2\mu_R}{\hbar^2K_0}
  \int_{r_p<r_p^{\rm ab}}  d\bfm{r}_p
  \int_{r_n<r_n^{\rm ab}}  d\bfm{r}_n
  \bar{\Psi}^*(\bfm{r}_p,\bfm{r}_n)
  \{ W_p(\bfm{r}_p)+W_n(\bfm{r}_n) \} \bar{\Psi}(\bfm{r}_p,\bfm{r}_n),
\label{eq-CFX}
\end{eqnarray}
\begin{eqnarray}
\sigma_{\rm IF}^{(p)}=
  \frac{-2\mu_R}{\hbar^2K_0}
  \int_{r_p<r_p^{\rm ab}}  d\bfm{r}_p
  \int_{r_n>r_n^{\rm ab}}  d\bfm{r}_n
  \bar{\Psi}^*(\bfm{r}_p,\bfm{r}_n)
  W_p(\bfm{r}_p) \bar{\Psi}(\bfm{r}_p,\bfm{r}_n),
\label{eq-IFXp}
\end{eqnarray}
\begin{eqnarray}
\sigma_{\rm IF}^{(n)}=
  \frac{-2\mu_R}{\hbar^2K_0}
  \int_{r_p>r_p^{\rm ab}}  d\bfm{r}_p
  \int_{r_n<r_n^{\rm ab}}  d\bfm{r}_n
  \bar{\Psi}^*(\bfm{r}_p,\bfm{r}_n)
  W_n(\bfm{r}_n) \bar{\Psi}(\bfm{r}_p,\bfm{r}_n).
\label{eq-IFXn}
\end{eqnarray}
The total wave function $\bar{\Psi}(\bfm{r}_p,\bfm{r}_n)$ is
obtained from $\Psi(\bfm{R},\bfm{r})$, which is given by
the CDCC calculation, by the straightforward transformation of the
variables.

We remark here that the above expressions of the three components of the
total fusion cross section are obtained by properly considering the
physics condition on the absorption corresponding to each process as mentioned
above.
On the other hand, following the definition of the incomplete fusion cross
sections of Refs.~\citen{Diaz1} and \citen{Iijima},
$\sigma_{\rm IF}^{(c)}$ ($c=p$ or $n$)
could be expressed by
\begin{eqnarray}
\sigma_{\rm IF,prev}^{(c)}=
  \frac{-2\mu_R}{\hbar^2K_0}
  \sum_{JM} \sum_{i \neq 0}^{i_{\rm max}}
  \left\langle
  \left[\chi_i(\bfm{R}) \otimes \hat{\Phi}_i(\bfm{r})\right]_{JM}
  \bigg|
  W_c
  \bigg|
  \left[\chi_i(\bfm{R}) \otimes \hat{\Phi}_i(\bfm{r})\right]_{JM}
  \right\rangle,
\label{eq-IFXprev}
\end{eqnarray}
i.e., the integration was done in the entire regions of
($\bfm{R}$, $\bfm{r}$), with taking only the wave function in
the breakup channels.
The expression of Eq.~(\ref{eq-IFXprev}) is unphysical, because
(i) breakup channels can contribute not only the incomplete but also
complete fusion processes, (ii) a possible contribution from
the elastic channel is naively disregarded, and
(iii) couplings between the channels of the three-body system,
which have been included in the calculation of $\Psi(\bfm{R},\bfm{r})$,
are neglected in the evaluation of $\sigma_{\rm IF,prev}^{(c)}$;
the expression of the complete fusion given in Ref.~\citen{Diaz1},
$\sigma_{\rm CF,prev}$, has similar issues.

We apply the new method for calculating complete and incomplete fusion
cross sections to the deuteron induced reactions on the $^7$Li
target for 10 MeV $\le E_d^{\rm L} \le$ 50 MeV, where $E_d^{\rm L}$
is the deuteron incident energy in the laboratory system. We use
the CDCC codes {\sc cdcdeu} and {\sc hicadeu} \cite{code}
to obtain the CDCC wave function $\Psi$ with assuming intrinsic
spins of $p$, $n$, and $^7$Li to be zero.
As for $V_{pn}$, we adopt the Ohmura
potential \cite{Ohmura} that reproduces the deuteron energy in the
ground state, i.e., $\varepsilon_0=-2.23$ MeV.
In the calculation of the $p$-$n$ discretized continuum states,
we include the s- and d-wave states; the maximum relative wave number
$k_{\rm max}$ is determined by
\[
k_{\rm max}=\frac{1}{\hbar}\sqrt{2\mu_r(E_d^{\rm CM}-|\varepsilon_0|)}
\]
with $E_d^{\rm CM}$ being the $d$-$^7$Li relative energy.
The $p$-$n$ continuum state is divided into 4 bin states,
for each of the s- and d-waves.
As for the $p$-$^7$Li and $n$-$^7$Li optical potentials,
we use the parameter sets in Ref.~\citen{Ye} except that
the spin-orbit terms are neglected in this study.

In the present formalism, the absorption radius $r_c^{\rm ab}$
is assumed to be a free parameter. In fact, it is found that
the results of the incomplete fusion cross sections calculated with
$r_c^{\rm ab}=4$ and 5 fm differ from each other by about 30\%.
Therefore, in this study, we determine $r_c^{\rm ab}$ at
$E_d^{\rm L}=40$ MeV so that the $\sigma_{\rm IF}^{(p)}$
agrees with the result of the Glauber model calculation; the latter,
added by the elastic breakup contribution calculated with CDCC,
reproduces the experimental DDX data very well at the same incident
energy.
The absorption radius thus determined is 4.0 fm, which is
used for both $p$ and $n$ in all calculations in this study.

\begin{figure}[!b]
\centerline{\includegraphics[width=70mm,keepaspectratio]{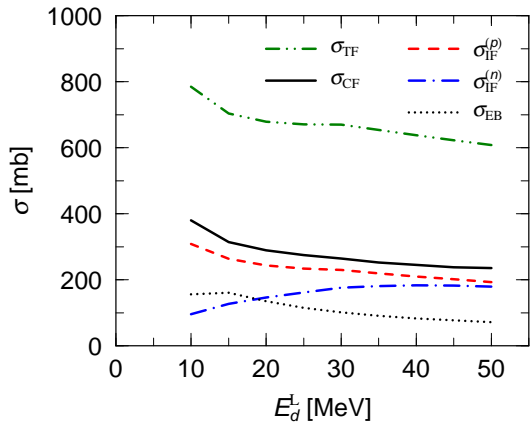}}
\caption{\label{fig-FX}
(Color online) Results of
$\sigma_{\rm TF}$ (dash-double-dotted line),
$\sigma_{\rm CF}$ (solid line), $\sigma_{\rm IF}^{(p)}$ (dashed line),
and $\sigma_{\rm IF}^{(n)}$ (dash-dotted line) for the
deuteron induced reaction on $^7$Li as a function of the incident
energy $E_d^{\rm L}$.
The dotted line represents the elastic breakup cross sections
calculated with CDCC.
}
\end{figure}
Figure \ref{fig-FX} shows the results calculated with the new method;
the dash-double-dotted, solid, dashed, and dash-dotted lines represent
$\sigma_{\rm TF}$,
$\sigma_{\rm CF}$, $\sigma_{\rm IF}^{(p)}$, and $\sigma_{\rm IF}^{(n)}$,
respectively.
One sees that $\sigma_{\rm CF}$ has the largest contribution to
$\sigma_{\rm TF}$ in the energy region of our interest.
Another important feature is that
the energy dependence of $\sigma_{\rm IF}^{(p)}$ is
significantly different from that of $\sigma_{\rm IF}^{(n)}$;
at $E_d^{\rm L}=10$ MeV, $\sigma_{\rm IF}^{(p)}$ is
three times as large as $\sigma_{\rm IF}^{(n)}$.
This is due to the difference in the energy dependence of
$W_p$ and $W_n$, i.e., $W_p$ ($W_n$) at low energy is more
(less) absorptive than that at around 40 MeV.
The elastic breakup cross section $\sigma_{\rm EB}$ is also shown
by the dotted-line in Fig.~\ref{fig-FX}.
The total neutron emission cross section, except for those by
the compound and preequilibrium processes, can be evaluated
as the sum of $\sigma_{\rm IF}^{(p)}$ and $\sigma_{\rm EB}$.
It is found that contribution of $\sigma_{\rm EB}$
is much smaller than that of $\sigma_{\rm IF}^{(p)}$,
which is consistent with the conclusion of Ref.~\citen{Ye3}.
On the other hand, if we consider the proton emission cross section
below 20 MeV, which consists of $\sigma_{\rm IF}^{(n)}$ and
$\sigma_{\rm EB}$, the two contributions are comparable.

\begin{figure}[!t]
\centerline{
\includegraphics[width=70mm,keepaspectratio]{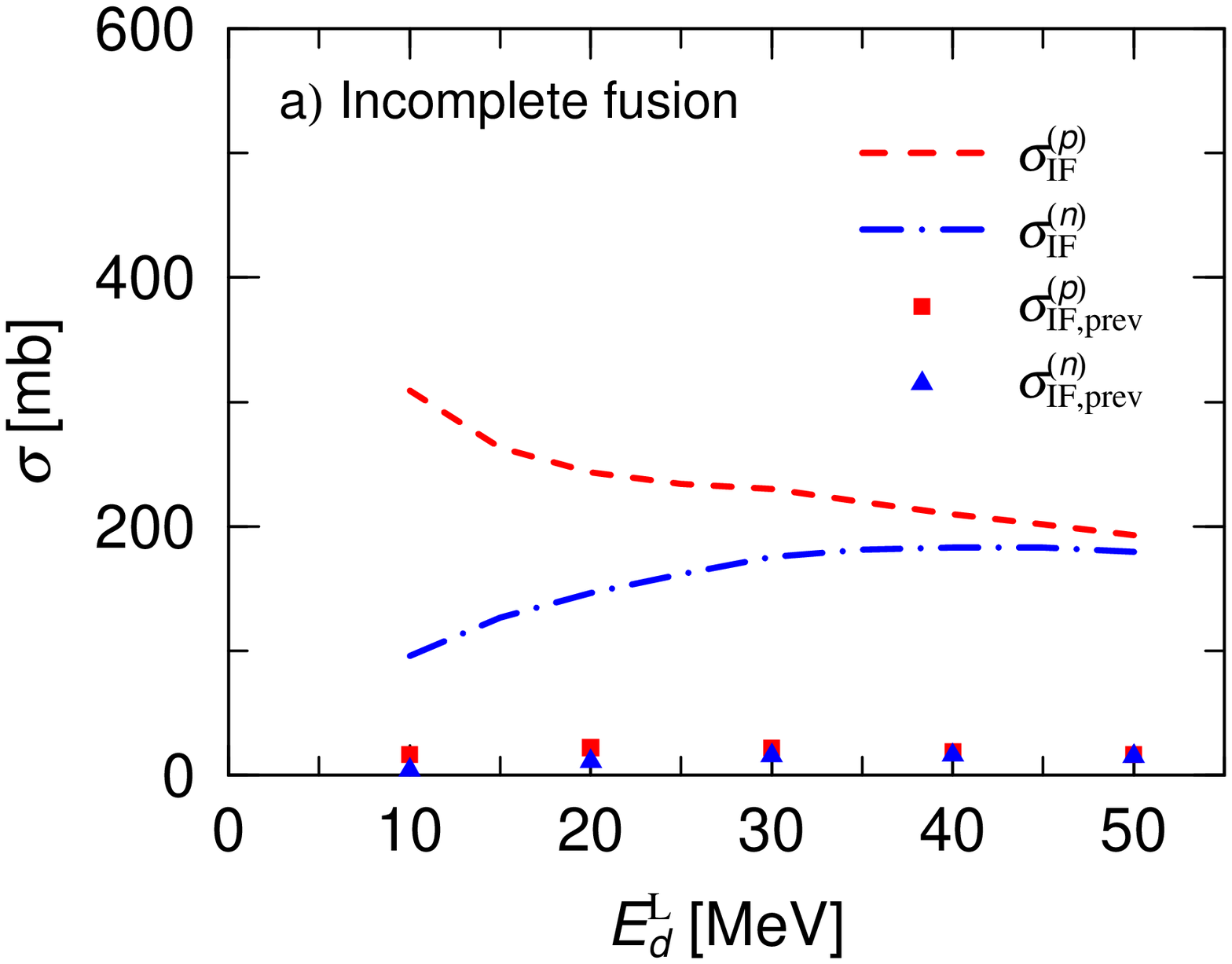}
\includegraphics[width=70mm,keepaspectratio]{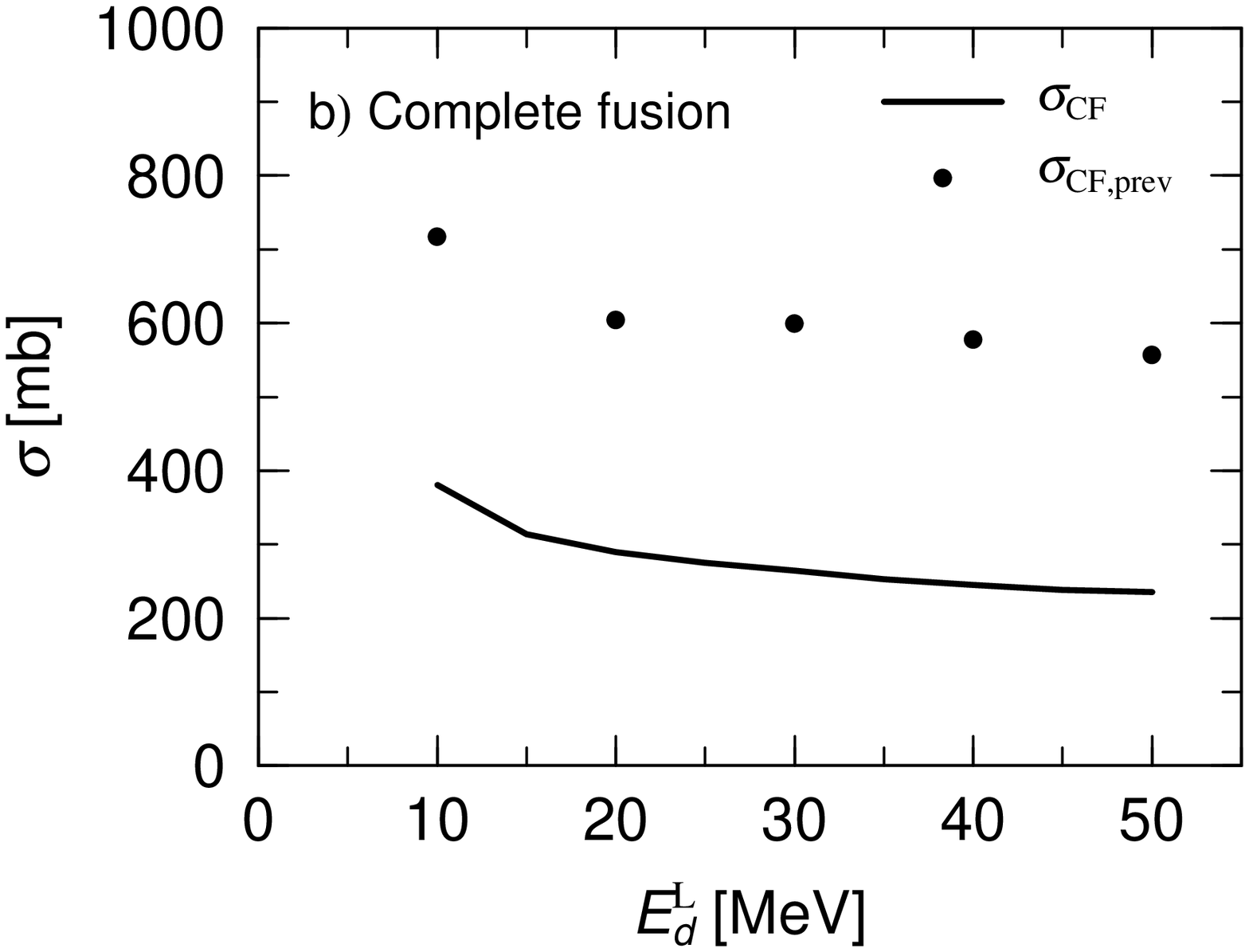}
}
\caption{\label{fig-FXwPrev}
(Color online)
a) $\sigma_{\rm IF}^{(p)}$ (dashed line)
and $\sigma_{\rm IF}^{(n)}$ (dash-dotted line) calculated with the present
method, compared with $\sigma_{\rm IF,prev}^{(p)}$ (squares)
and $\sigma_{\rm IF,prev}^{(n)}$ (triangles) following the previous
definition given in Refs.~\citen{Diaz1} and \citen{Iijima}.
b) Comparison between the results of the complete fusion cross sections
calculated with the present (solid line) and previous (dots) method.
}
\end{figure}
Next we show in Fig.~\ref{fig-FXwPrev}
the incomplete and complete fusion cross sections calculated with the
previous expressions in Refs.~\citen{Diaz1} and \citen{Iijima},
compared with the results of the present study.
The squares and triangles in the left panel show, respectively,
$\sigma_{\rm IF,prev}^{(p)}$ and $\sigma_{\rm IF,prev}^{(n)}$,
and the dots in the right panel show $\sigma_{\rm CF,prev}$.
The lines shown in the panels are the same as in Fig.~\ref{fig-FX}.
One sees clearly that the previous prescription gives much smaller
(larger) incomplete (complete) cross sections than those obtained
by the present calculation.
Our result of $\sigma_{\rm IF}^{(p)}$ at 40 MeV, by definition,
can be interpreted as an experimental value. Thus, the prescription
given in Refs.~\citen{Diaz1} and \citen{Iijima} do not work
at all at least for the inclusive $^7$Li($d,n$) reaction concerned.
In other words, Fig.~\ref{fig-FXwPrev} clearly
shows the importance of including elastic channel in the
evaluation of $\sigma_{\rm IF}^{(c)}$ ($c=p$ or $n$)
as in Eqs.~(\ref{eq-IFXp}) and (\ref{eq-IFXn}). Similarly,
inclusion of the breakup channels in the calculation of
$\sigma_{\rm CF}$ is also important.

\begin{figure}[!t]
\centerline{
\includegraphics[width=60mm,keepaspectratio]{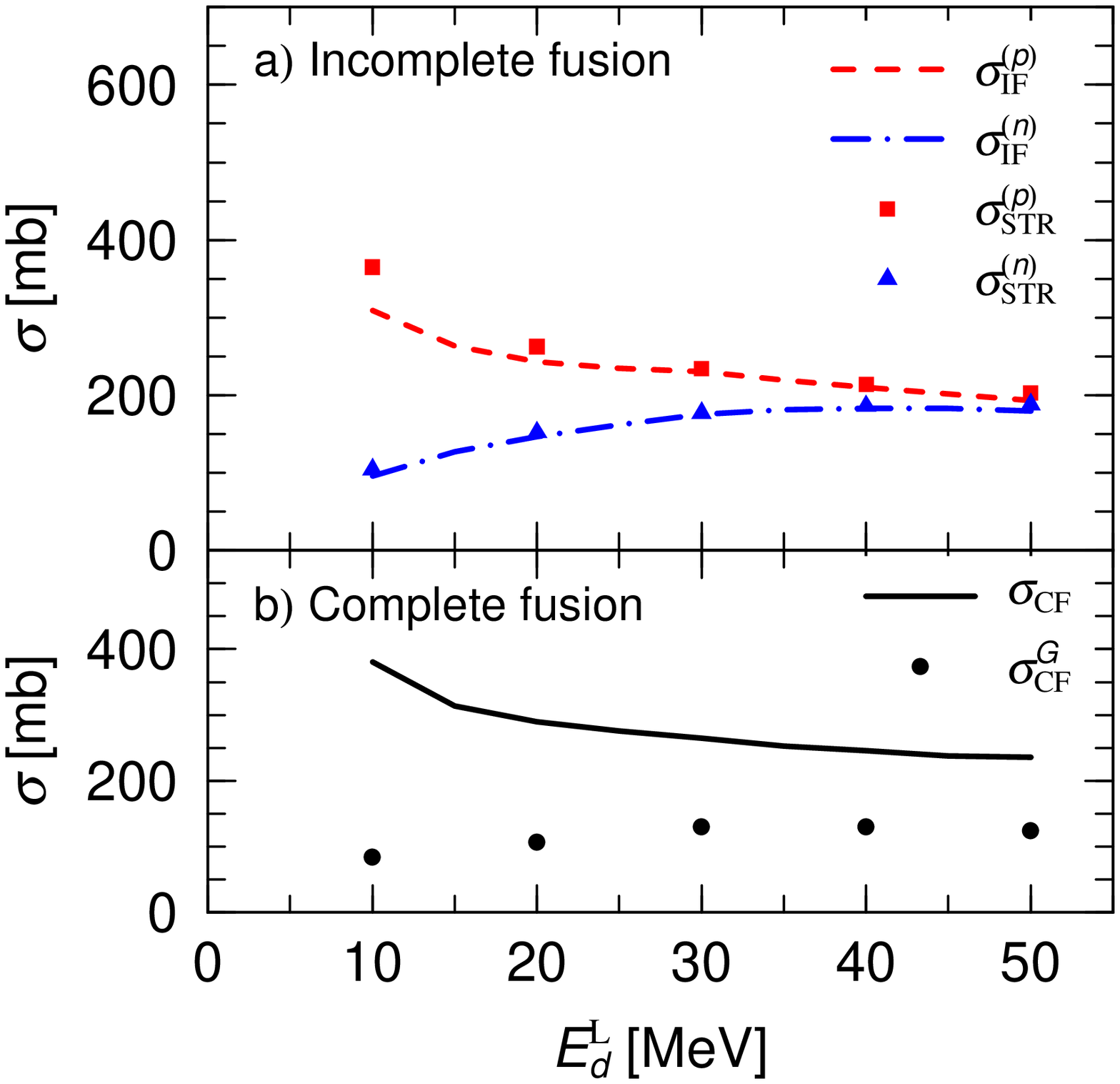}
\includegraphics[width=60mm,keepaspectratio]{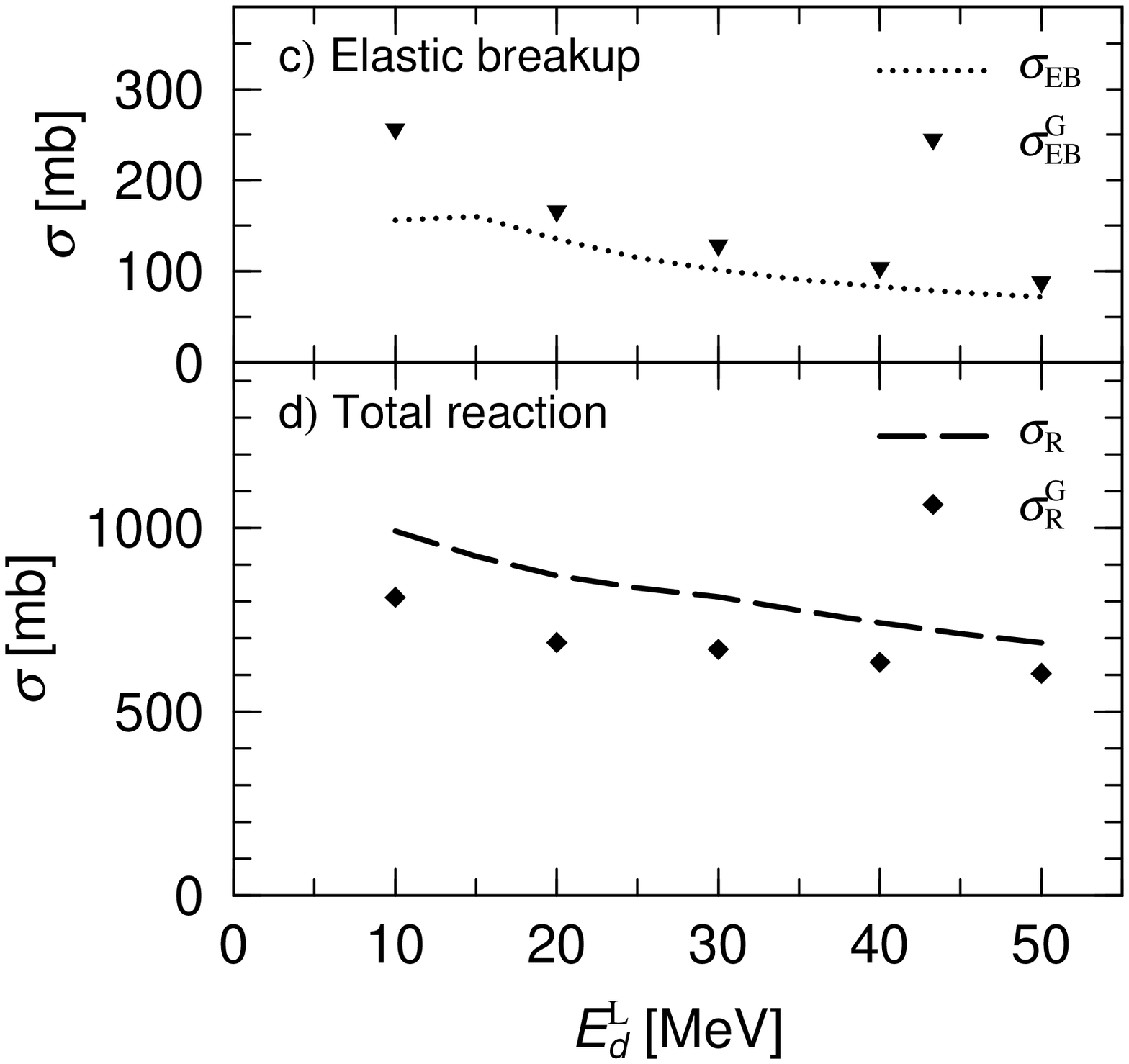}
}
\caption{\label{fig-FXwG}
(Color online)
a) Comparison of $\sigma_{\rm IF}^{(p)}$ (short-dashed line)
and $\sigma_{\rm IF}^{(n)}$ (dash-dotted line) obtained by the
present calculation with $\sigma_{\rm STR}^{(p)}$ (squares)
and $\sigma_{\rm STR}^{(n)}$ (triangles)
by the Glauber model.~\cite{Ye3} \
b) Complete fusion cross sections calculated with
the Glauber model (dots) and the present method
(solid line).
c) The elastic breakup cross section calculated with CDCC
(dotted line) compared with that with the Glauber model
(inverse triangles).
d) The total reaction cross section obtained by
CDCC (long-dashed line) is compared with that by
the Glauber model (diamonds).
}
\end{figure}

As mentioned above, in Ref.~\citen{Ye3}, the contribution
of the proton stripping process to the inclusive ($d,n$)
cross section at 40 MeV was shown to be described very well
by the Glauber model. It is thus interesting whether the
Glauber model calculation works or not at even lower
energies.
For this purpose, in the left-upper panel in Fig.~\ref{fig-FXwG},
we compare $\sigma_{\rm IF}^{(p)}$ (short-dashed line) calculated
with the present method, with $\sigma_{\rm STR}^{(p)}$ (squares)
obtained by the Glauber model calculation.\cite{Ye3} \
Surprisingly, the two results agree well each other not only
above 40 MeV but also at low energies down to 10 MeV.
This is also the case with the neutron stripping process;
one sees the good agreement between
$\sigma_{\rm IF}^{(n)}$ (dash-dotted line) and
$\sigma_{\rm STR}^{(n)}$ (triangles).
Note that we use the absorption radius of 4.0 fm determined
at 40 MeV in all the calculation as mentioned above.
Thus, we conclude that the Glauber model calculation for
the incomplete fusion cross sections is expected to work
even at lower energies down to 10 MeV.
On the other hand, as shown in the left-lower panel,
the results of the complete fusion cross section
$\sigma_{\rm CF}^{\rm G}$ obtained by the Glauber model (dots)
\cite{Watanabe} significantly deviate
from those obtained by the present study, i.e.,
$\sigma_{\rm CF}$ (solid line);
even the energy dependence is different.

A possible reason for the success of the Glauber model
in describing stripping processes is, as discussed in Ref.~\citen{Ye3},
that the contribution from the nuclear surface region is dominant,
where the depth of the optical potential is so shallow
that the Glauber model works well.
On the other hand, since the complete fusion process takes place
in the nuclear interior, the Glauber model does not work even at
50 MeV.
It is numerically confirmed that when we make artificially
the nucleon-$^7$Li optical potential shallow,
the difference shown in the left-lower panel becomes small,
while the features of the incomplete fusion cross sections
(left-upper panel) have no changes.

In the right-upper panel, we show the results of the
elastic breakup cross section calculated with CDCC (dotted line)
and the Glauber model (inverse triangles). The difference
between the two is quite small above 10 MeV, which is found
to be mainly due to the adiabatic approximation used in the
Glauber model. The quite big difference at 10 MeV will
come from the invalidity of the eikonal approximation
also assumed in the Glauber model.
The total reaction cross section, which is the sum of the
total fusion cross section and the elastic breakup one,
calculated with CDCC (the Glauber model) is shown
by the long-dashed line (diamonds) in the
right-lower panel.
The main part of the difference of the two
comes from that in the complete fusion cross sections.

In summary, we propose a new method for evaluating the complete
and incomplete fusion cross sections separately by means of CDCC.
The separation of the two is carried out by the physics
condition on the absorption for each fusion process.
The absorption radius included in the present formalism is
determined using the result of the proton stripping
cross section for the $^7$Li($d,n$) reaction at 40 MeV
calculated by the Glauber model, which
was shown to reproduce the corresponding experimental value.
The new method is applied to the $^7$Li($d,n$) reaction
from 10 to 50 MeV.
The complete
fusion cross section is found to have the largest contribution
to the total fusion cross section.
The ($d,p$) and ($d,n$) incomplete fusion cross sections
show quite different energy dependence, because of that
in the imaginary parts of the $p$-$^7$Li and $n$-$^7$Li
optical potentials.
It is found that in the all energy region considered,
the ($d,p$) and ($d,n$) incomplete fusion cross sections
obtained by the Glauber model agree well with those
obtained by the present calculation with CDCC.
On the other hand, the two model calculations give
significantly different results of the complete fusion
cross section, even at 50 MeV.
The complete and incomplete fusion cross sections
obtained by the previous method of Refs.~\citen{Diaz1}
and \citen{Iijima} are found to be inaccurate.
Extension of the present framework to calculate the DDX
is a very important future work.
A method to divide the complete and incomplete
processes unambiguously, i.e., without the absorption radius,
will also be desirable.

\vspace{3mm}

We would like to thank Y.~Watanabe and T.~Ye for fruitful discussions and
providing the numerical results.
We also acknowledge helpful discussions with K.~Hagino.
SH thanks Y.~Aoki for valuable discussions.

%

\end{document}